# On how the Cyberspace arose to fulfill theoretical physicists' needs and eventually changed the World: Personal recallings and a practitioner's perspective[1]

Emilio Elizalde

National Higher Research Council of Spain
Instituto de Ciencias del Espacio (ICE/CSIC and IEEC)
Campus Universitat Autònoma de Barcelona (Barcelona) Spain[2]

**Outline:**
   Abstract & Key words
1. Introduction
2. A most humble aim: Producing professionally-looking preprints and sending their encoded files by using computers
3. Paul Ginsparg and the ArXiv concept
4. The World Wide Web
5. Fulfilling the needs of theoretical physicists
6. Once more the same discourse, but from a difeerent perspective
7. The uncountably many present uses of the Cyberspace
8. The dangers of the Cyberspace
9. The future of the Cyberspace
10. Conclusion
11. References
12. Author's short biosketch

**Abstract**
The very humble origins of the Cyberspace, and all the related developments that smoothly conspired and converged towards this concept, making its emergence possible, as the personal computer, TEX and LATEX, the Fax, the internet, the cellphone, and the World Wide Web, are discussed, always from a personal perspective. A separate, comprehensive explanation of the reasons for the appearance and subsequent evolution of each of these different phenomena, with explicit assessments and a future prospective of the main subject, is presented.

---

# 1. Introduction

Let me start from the beginning. I am a theoretical physicist and a cosmologist, as well as a mathematician [Elizalde 2016p]. My main research fields are zeta function regularization [Elizalde 2012, Elizalde et al 1994], dark energy issues [Elizalde et al 2004, Elizalde et al 2008], mainly in relation with quantum vacuum fluctuations [Bordag et al 1996], and modified gravity [Elizalde et al 2003]. In 1982, during my second visit in Hamburg (the II. Institut für Theoretische Physik and the Deutsches Elektronen-Synchrotron (DESY), as a Humboldt Fellow [AvHumboldtF 2016, AvHumboldtE 2016], I came to know about the TEX project, a typesetting system developed by Donald Knuth [Knuth 1969], and existing since 1978. It was a wonderful, very ambitious project. A couple of years later, back at the Department of Theoretical Physics of Barcelona University, as an Assistant Professor, I was chosen to be the responsible for the preprint interchange with other universities worldwide. Sure, this had always been a very important issue for theoretical physicists anywhere, right then and also in the past. The reason was the following. When one would send a paper for publication to a specialized journal, one would always have to wait for several months (sometimes one or two years) before the paper managed to make its way to a publication (even today this is a serious issue [Powell 2016]). In a very quickly developing field, as the one in question, this was at the time a really dramatic drawback. In other words, the importance of a rapid dissemination of our work, under the form of the so-called preprint [Preprint 2016], was paramount. But the whole procedure was extremely cumbersome, as I could appreciate (suffer is a better word), first hand, during the period that I was in charge of this business for my Department. I will now try to describe what was going on.

To start with, we had to do a careful selection of a certain number of the best Departments of Physics in the World [Universityrankings 2016], being the very first step to decide on this precise number; what depended on the money available to carry out the full task. At the epoch I was in charge, we selected in Barcelona over 200 universities (250 was soon to become our all-time record), with annual revision of the list. We were bound to choose the most interesting places where to send the preprints, both because we assured in this way that our work would get to be known by the top specialists in our research fields and-equally or even more important-because there was an (unwritten) law of reciprocity, by which, on doing so, we would also get the preprints issued by those departments where the best of our colleagues were working at that time. I can assure you, choosing the list was not always that simple [Topcitestheophys 2014].

The second step, after completing some valuable, original research project (this was actually our real duty, of course, aside from giving lectures), was to type our preprints using our precious typewriter (sometimes there was a line to be kept for that), taking care to allow enough space for the special characters, formulas and equations, which we had to fill in later, always by hand. If something got wrong during the typing, or if we discovered some mistake, we had to start over, sometimes right from the beginning! Each month or two I would collect all preprints written in our group and, with the help of a couple of post-docs we would make the, say 220 copies of each of the say 8 papers;

what then meant classifying the pages, stapling each copy together, writing the envelopes with the 220 addresses, filling each one of the envelopes with the 8 copies, closing and sealing the 220 envelopes, bringing the whole package to the post and, finally, paying good money for the expedition of the package.

Even now, by just recalling and describing to you in detail the whole process we had to go through I get again pretty tired. The whole thing had a considerable cost, both in terms of personal effort and money (and also ecological! although this concept just did not exist right then). You should keep in mind, and perhaps read once more, the several paragraphs above in order to better grasp the enormous importance it had for us the developments that were about going to take place, almost simultaneously and on several parallel, complementary levels, and which eventually conspired together to give birth to the most important social revolution, at planetary scale, in human history.

## 2. A most humble aim: Producing professionally-looking preprints and sending their encoded files by using computers

The first remarkable development, as already advanced in the Introduction, was the appearance of TEX [Knuth 1969], and a few years later, in the middle 80's, of LATEX (Leslie Lamport's TEX) [Lamport 1986], the poor brother of TEX but much more simple to use and thus more appropriate for the average theoretical physicist (or mathematician). LATEX (even more, TEX) was (and still is) an incredibly powerful tool. It permitted you to write absolutely perfect, professionally looking papers, allowing anybody to produce high-quality articles and books using fairly minimal effort. And that under the form of a binary file, which could be sent away without the least distortion, not even of a single bit (lossless, as we would say today [Infotheor 2016]): it would yield exactly the same final result on all kind of computers, at any place in the world and point in time.

Needless to say, typewriters had in the meanwhile (in just very few years) become fully obsolete with the advent of personal computers: Atari, Altair, Spectrum, Amstrad, IBM, to name only a few of the most significant ones, in terms of popularity [Knight 2014]. The marvelous and very modern IBM ball typewriters I had used to type my papers when I had been at DESY for the first time (1979-80) were suddenly of no use any more, at least for us, theoretical physicists (although they were still employed, for a while, for writing down ordinary forms or filling up offcial documents).

But this immediately leads us to the second extraordinary development (in fact, the first one in order of importance, even more with regard to the title of the present book). To wit, the LATEX file encoding all information of the paper (or book, given the case) had to be transmitted, e.g. distributed, as with previous, paper preprints was done in the way I have described before. However, this was not going to take place, now, by using the pony express or any other postal services anymore. In no way by sending again, inside of an envelope, the magnetic tapes or discs or any kind of material support containing the files. The groundbreaking idea was now to connect computers to computers and transmit the files immaterially, using this connection: the internet.

A really nice definition from the Wikipedia:

> "the Internet is the global system of interconnected computer networks
> that use the Internet protocol suite TCP/IP [Tcpip 2016] to link billions
> of devices worldwide. It is a network of networks that consists of
> millions of private, public, academic, business, and government networks
> of local to global scope, linked by a broad array of electronic, wireless,
> and optical networking technologies." [Internet 2016].

In the middle eighties of the past century, aside from an extensive military network already in place in the USA, the Arpanet (existing since the 1960's, and which was in fact the origin of the concept of the present internet itself), `the international internet' consisted in a connection of just *three* computers: one in Los Alamos National Laboratory (LANL), USA [Lanl 2016], one in the European Organization for Nuclear Research (CERN, in Geneva, Switzerland) [Cern 2016], and one in DESY (Hamburg, Germany) [Desy 2016]. Actually, these three institutions had already been playing, for many years, a very important role for us all, theoretical physicists: they maintained archives where all preprints they received where classified by subject, given a number, and properly marked with a reception date. Those were considered by everybody to be very reliable `oficial' archives; for instance, in order to establish priority of a certain result or discovery. Of course this had been always established by the date of the corresponding publication, finally, in an international journal. But in such a fast developing area no one would wait until publication, to establish such priority: a clever idea or result could be stolen (and some certainly are [Southwick 2012]), or at the very least much delayed, during the long revision process that led to publication.

With the use of these archives, priority was established, in fact, since the very moment when your work got registered in any of those repositories (better in more than one!) under the form of an archived preprint with a reception date. Just because these archives were reliable, public, independent and fully transparent to the whole international community. This issue was actually of paramount important, in particular, for a young scientist as I myself was at that time. One always had the highest expectations and believed that the work just finished was to become a masterpiece, or, at the very least, a very much influential paper.

With the new advances, Los Alamos, CERN, and DESY could actively share their collections of preprints and start to work together. The first step was taken to provide open access, from the distance, to the electronic preprint collections by means of the internet computer connections already mentioned. Thus, in DESY there was a list where you would enter your name, and a date and time slot to be filled, which procured you half an hour of use of the computer, which you then would spend by looking from Hamburg's computer at the preprint lists at CERN and Los Alamos. You then could write down the references you were interested in and ask later the authors of the papers to send a copy to your address. It took still some time before we were able to retrieve the corresponding file through the computer connection (for it was quite slow and of poor bandwidth). Needless to say, all enquires where then done by entering your computer orders line by line, being UNIX [Unix 2016] the computer language, and using the whole screen at a time (the Windows concept had still to appear).

Now a revealing anecdote. That same summer my two sons, both music students, who were staying with me and my wife at the DESY Hostel for the period, needed to practice on a piano during their long vacations. At DESY the only piano available was the one in the main Auditorium. It is still there, at the same right corner of the stage, and that after 30 years, as I could check last month [Desy 2016b] when I visited DESY again, to take part in the Annual Conference and in the Colloquium *Local Quantum Physics and beyond*, in memoriam Rudolf Haag (by the way, a renowned professor, who had been my host during my stay in Hamburg as a Humboldt Fellow, in the 1980's). Curiously enough, the procedure I had to follow in order to reserve the piano for my sons to play, and the one for reserving the computer for me to use, as explained, were exactly the same! I would bet that, with the piano, you still have to go right now through the same procedure (the only possible improvement being that you will now enter your appointment at the piano list conveniently using your cellular). This makes me ponder how enormously things have changed, during these same 30 years, concerning the uses and possibilities of computers. In particular, in relation with the issue discussed in this paragraph: entering a list to reserve computer time to connect somewhere (...?) seems a ridiculous concept, of the Stone Age! It is amazing!

## 3. Paul Ginsparg and the ArXiv concept

Another important step was going to be taken soon. The ArXiv, a repository of electronic preprints or *e-prints*─which was the name soon given to the LATEX files I was talking about─was put in place on August 14, 1991. As clearly explained in the Wikipedia [WikiArxiv 2016], it was made possible by the lowbandwidth TEX file format, which allowed scientific papers to be easily transmitted over the incipient internet. In fact, according to the Wikipedia, this had been started a bit before, around 1990, by Joanne Cohn, who had begun e-mailing physics preprints to colleagues as TEX files, but the number of papers being sent soon surpassed the very limited capacity of mailboxes. Paul Ginsparg [Ginsparg 2016], a theoretical physicist who had begun as a junior fellow at Harvard and had moved as a member of staff to LANL, quickly understood the need for a capable central storage device, and in August 1991 he finally materialized the idea to install a central repository mailbox to be accessed from all computers with an internet connection. He was very proud that he could store so many preprints in his own, not so big computer, and I can certify that the service he provided to the international community of theoretical physicists, for the first few years at least, with such limited resources, was incredible.

Important additional access modes were added soon, namely the File Transfer Protocol (FTP) in 1991, Gopher in 1992, and in 1993 the World Wide Web (WWW). I distinctly remember all these steps, as very welcome additions to the process, constantly improving preprint production, file transmission, and access to the repository. The ArXiv began as a theoretical physics archive only: the LANL preprint archive, its domain name being xxx.lanl.gov. Soon other disciplines of physics were added, and then astronomy, mathematics, computer science, nonlinear science, quantitative biology, statistics and, lastly, quantitative finance. Ginsparg moved to Cornell University in 1999 and he changed the name of the repository to arXiv.org. Presently, it is hosted by Cornell, but it has 8 mirrors scattered around the world. In 2008 the ArXiv reached the half-million article mark, and the one million milestone in 2014. The submission rate approaches the 10,000 papers per month, right now. Not just the volume, but the quality and importance of the ArXiv have been increasing constantly and, in a way, the whole process has culminated in the most relevant publishing movement of our days, known as Open Access [Openaccess 2016].

At a point, the registration of a preprint in the ArXiv [Arxiv 2016] has been given equal status to that of a paper regularly published in a most prestigious specialized journal. For one, let me just recall the famous case of the Russian mathematician Grigoriy Perelman [Perelman 2014], who proved the Poincaré conjecture [ClayPoinc 2016a], the first (and the only one till now) of the Seven Millennium Problems [Milleniump 2016] to have been solved. Perelman issued his solution in a series of three e-prints that appeared in the ArXiv between Nov. 2002 and Jul. 2003 [Perelman 2002; Perelman 20023a; Perelman 2003b], but which he refused to send to a journal for publication, with the argument that this was fully unnecessary, given the universality of the ArXiv. He went on saying that everybody in the World could check if his results were right-of what he was completely sure and was, in fact, definitely proven to be the case. Perelman's proof of the Poincaré conjecture is considered to be one of the most important achievements of the past Century.

Although, in order to meet the criteria to receive the one-million-dollar Clay Millennium Prize [ClayPoinc 2016b] for this discovery, it was compulsory that the proof would be published in a regular journal, the Clay Foundation gave finally up, and offcially acknowledged (for the first time ever) that Perelman's e-prints registered in the ArXiv were fully equivalent to regular journal publications (for the reasons already given). This was the culmination of the idea, I had advanced in a previous section, that priority of a scientific result is now fully, legally established in terms of its acceptance and registration as an e-print in the ArXiv.

The above case has also prompted many other colleagues, some of them quite distinguished, to do the same as Perelman, at least from time to time: the number of important results that have only been issued as e-prints, never having been published in a regular journal, is constantly increasing-and a number of them have accumulated record numbers of citations [Lariviere et al 2013].

## 4. The World Wide Web

Scholars generally agree, as well described in the Wikipedia, that a turning point for the WWW began with the introduction of the Mosaic web browser [Mosaic 2016] in 1993, a graphical browser that had been insistently demanded by our community of theoreticians in order to deal with figures and graphs (I will discuss this issue in much more detail later). Before the appearance of Mosaic, graphics could not be easily incorporated in web pages, together with text. As a consequence, the Web was not so popular as other protocols, as e.g. Gopher. The graphical user interface of Mosaic was decisive in making of the Web, by far, the most popular internet protocol. In October 1994, the physicist Tim Berners-Lee founded the World Wide Web Consortium (W3C) at the Massachusetts Institute of Technology (MIT), what was the culmination of this process. But let us proceed slowly and recall in detail the whole story.

Soon after graduation, Berners-Lee had become a telecommunications engineer and after working for the industry he had been a fellow at CERN, in Geneva (Switzerland), for ten years, where he first became acquainted with the needs of the huge community of physicists working there and then devoted his best efforts to satisfy them. We should not forget that the CERN, already mentioned in previous sections, is now the greatest temple of Theoretical and High Energy Physics in the World. It has the most powerful particle accelerator, by far, namely the Large Hadron Collider (LHC). Experiments in the LHC have allowed to reproduce and study in all detail the same physical conditions of our Universe when its age was just of some three femtoseconds, that is, when just $3\times 10^{15}$ s had elapsed since the Big Bang singularity.

CERN was already at the time the largest Internet node in Europe, and Berners-Lee clearly saw an opportunity there, as he personally explains:
> "I just had to take the hypertext idea and connect it to the transmission control protocol (TCP) and domain name system (DNS) ideas and ta-da! the World Wide Web."… "Creating the web was really an act of desperation, because the situation without it was very dificult when I was working at CERN later. Most of the technology involved in the web, like the hypertext, like the Internet, multifont text objects, had all been designed already. I just had to put them together. It was a step of generalizing, going to a higher level of abstraction, thinking about all the documentation systems out there as being possibly part of a larger imaginary documentation system." [Berners-Lee, 2016]

However, as advanced, the Web was actually founded when Berners-Lee had already left CERN, at another World leading institution: The Massachusetts Institute of Technology Laboratory for Computer Science (MIT/LCS) with support from the Defense Advanced Research Projects Agency (DARPA [Darpa 2016]), which had pioneered the Internet, as already explained above. A second site was founded just one year later, at the French laboratory Institut National de Recherche en Informatique et en Automatique (INRIA), with support from the European Commission Digital Information Society (DG InfSo, recently renamed DG Connect). A third continental site was then created at Keio University, in Japan, in 1996. Although in the middle 1990's the number of websites remained still relatively small, many of the best known sites were already functioning and providing popular services.

Probably the most important idea of Berners-Lee, while at CERN, was to join together the hypertext and the internet. In doing so, he developed three essential technologies, which sound very familiar now [Http 2016]: (i) a system of global, unique identifiers for all resources on the Web, first known as the Universal Document Identifier (UDI), and later specified as the Uniform Resource Locator (URL), on one hand, and the Uniform Resource Identifier (URI), on the other; (ii) the publishing language HyperText Markup Language (HTML); and (iii) the HyperText Transfer Protocol (HTTP).

The Web had a number of differences with respect to the other available hypertext systems, in that it required only unidirectional links, and not bidirectional ones, as the rest. This fact rendered it possible, for anybody, to establish a link to another computer without the need to have to wait for any action by the responsible of the second resource. It simplified very much the process of implementation of web servers and browsers. As soon as 30 April 1993, CERN officially announced, that the WWW would be free to anyone [Birth-web 2016]. This fact that the Web was non-proprietary was decisive, since it made possible to develop new servers and clients and to add extensions without having to ask for licenses all the time. As the use of the Gopher competing protocol was not free, this announcement immediately produced a decisive movement of clients towards the Web. Berners-Lee had made his brilliant idea freely available, with no patent to be filled and no royalties to be paid. The WWW Consortium [W3 2016] also decided, on its turn, that Web standards should be based on royalty-free technology, so that they could be easily adopted by everybody in the World. This fact contributed to the extremely quick and successful expansion of the Web in the whole World since the middle 1990's.

The impact of the Web on the world society has been enormous since then, but not so specially on theoretical physicists, mathematicians or theoreticians of any other kind, as I will now explain below. The first web site was opened at CERN, and had the address: info.cern.ch. The first web page http://info.cern.ch/hypertext/WWW/TheProject.html contained information on the WWW project. On Aug. 6, 1991 it was set online and, as the project was quickly improving, changes to the page were made each day. In a list of *"80 cultural moments that shaped the world,"* chosen by a panel of 25 eminent scientists, academics, writers and world leaders, the WWW has been ranked as number one of all them, for being

> "The fastest growing communications medium of all time, the Internet
> has changed the shape of modern life forever. We can connect with each
> other instantly, all over the world." [British Council 2014]

This development was absolutely unforeseen by the creators of the Web and by the responsible at CERN and MIT. No one could guess that by just trying to implement, to bring to its ultimate perfection, the idea of dealing with all aspects of scientific work production, sharing, cooperation and distribution, by then going further to include figures, images and even little movies, necessary to better describe some important scientific results, would be later used to satisfy so many different needs of all areas of knowledge and, far outside from that, of the common human being and of the universal society. And that it was going to fulfill dreams no one had even imagined to be able to dream in the late 1990's. In the following sections I will further elaborate around this issue.

## 5. Fulfilling the needs of theoretical physicists

When I now compare what I, as a theoretical cosmologist or mathematician, have gained from the recent web developments, say since day one of the present Millennium, my answer must be: fairly little. The needs of an average theoretician or high-energy physicist, which were at the very origin and conception of the humble initial proposal, which culminated in the Cyberspace, had been totally fulfilled by the end of the last Millennium, even before, probably. I will put a specific example in order to justify and properly explain this claim. During my career as a physicist, I have had a number of PhD students and I will now just compare their daily work and behavior, during the preparation and finalization of their respective thesis works and until they got a Post-Doc.

My firsts students, in the middle to late 1980's used typewriters for writing down the text part of their manuscripts, filling later formulas, equations and special characters by hand. They customarily attended international schools and conferences and were much aware of the latest results in their research subjects through preprints that they solicited by postcard, normally to the authors themselves or, alternatively, to the already mentioned repositories. At that time, if they were lucky, they could manage that some preprint was sent to them by Fax (what had some associated cost that I would cover through my research funds); this was indeed another precious possibility at the time, not mentioned before, which I will consider later in further detail, and in a more general context. Although they also consulted the preprints displayed on an expository shelf at our Department of Theoretical Physics (University of Barcelona, UB), arriving from Los Alamos, CERN, DESY and several dozen other universities (see Sect. 1).

The appearance of LATEX, which we got in 1987, was enthusiastically welcomed in our Department, but not so by other groups of the Physics Faculty, who remained quite indifferent or were unwilling to lose even the least amount of time in that matter. Actually my then student Enrique Gaztañaga and myself were the very first in the University of Barcelona (UB) to get the LATEX program; I clearly remember the day when we went to visit the group of Theoretical Physics of the Autonomous University of Barcelona (UAB)-which was the first group to buy the program in our local community-to get hold of a copy. Soon we started, with Enrique, to write preprints in LATEX; this was certainly a revolution and, needless to say, he wrote his PhD thesis in this code. Compiling each page took a couple of minutes at the time, but the final result was astonishingly professional. And you could improve the manuscript so easily, spotting the mistake and changing only this little part. Sure, you cannot experience such a wonderful feeling in all its depth if you did not live through this period, if you were not actually there. My student August Romeo was also born to theoretical physics at that time. Both August and Enrique soon mastered in LATEX and I learned a lot from them.

I have already described the fast evolution and improvements that took place immediately afterwards. In 1991 the ArXiv was born and there was no more need to send postcards in order to get interesting preprints, nor had we to wait for them to be displayed on our preprint shelf on Friday morning every week. Of course, some improvements continued to happen in the 1990's, as in the searching, classification, speed in sending and getting e-prints, compilation got much faster, etc. But from the end of the 1990's onward the advances, as what concerns our field of research, have been rather cosmetic or superficial.

When I now do compare how my students from the late 1990's and my present students work, the differences are no more that remarkable. They regularly access the ArXiv (now from any place, that is true, including their cellulars) to easily search for information about the subject of their work. They write papers in LATEX and send them also to the ArXiv, on their turn. However, I would bet that (at least on the average) present day theoretical physics students do not master LATEX to the same level as we, little experts, used to do in the old times. For us it has always remained something very, very special.

Today they have several other alternative possibilities. For one, the text processor Word [Word 2016], which was actually always there (in parallel to LATEX), to write ordinary documents, has become very powerful and allows now to insert beautiful equations, too. And, on the other hand, Word is also very useful in connection with clean and very professional presentations by means of Power Point [Powerpoint 2016]. Let me say, on passing, that Word and the rest of the Office programs represented for the administrators and the society in general, a parallel revolution as powerful and important (or even much more, according to the number of people influenced by it) as that of TEX for the scientists. My whole point is just that we, theoretical physicists, were the spark, the ones who ignited the fire, and, much more than that, the ones who created the whole firework castle, e.g., the language, the connections, the whole infrastructure that made it possible for the whole thing to explode and expand.

Actually, talking about the Office programs, this is another important issue I have not been dealing with before, namely that of the evolution of the way physics results are presented in seminars, lectures, and conferences of all kind. That is, the revolution that brought us from `transparencies', always written by hand, and `retro-projectors' —with the related thousandand-one problems of bulbs that kept fusing all the time-to the ubiquitous use of computers and Power Point (or Acrobat pdf [Acrobat Adobe 2016]) presentations with animations of all sort trying to captivate the audience at any price. There has been an enormous change in that direction but I would say this one is not of an essential nature, in the sense that the other changes here described have been.

To establish this point on firm grounds and with very few words, let me just mention a couple of examples. Recently, in the spectacular Starmus meeting in Tenerife [Starmus 2016] I could witness (and a dozen Nobel Prize Laureates participating there also) how Sir Roger Penrose gave his awaited for presentation [Penrose 2016]. He did it with the help of a computer and other modern means, of course, but it simply consisted of a bunch of projected pdf scans of ordinary handwritten pages. They looked exactly as in the old times! It was a big surprise for the whole audience. But this one is in no way the only such case I have seen in the last half a year (another colleague did the same in a scientific presentation at the Benasque Center of Theoretical Physics in September this year). As another, different example I may add that I still attend, from time to time, to interesting seminars given on the blackboard. And, moreover, it is remarkable that many colleagues often tend to think that these presentations are in fact deeper, more interesting, and that they even captivate the attention from the audience better than many up to date Power Point fashionable flashes [Powerpoint 2016].[3]

Back to the issue under discussion above, my present students still study with the same reference e-books, and also with ordinary books, of course (sometimes coming from pirate sources), exactly as in the last decade of the past Millennium. Maybe the only significant difference is the one just mentioned by passing, that access to information about their working interests has now become much faster, space and timely ubiquitous. And also the existence now of the Wikipedia (not of much use for doing hard and original research work, actually), and of other internet resources, as the extraordinary MIT on-line lectures [Lewin 2016], and those from many other top universities and research labs. But neither of those are really fundamental improvements for the work of a theoretical physicist or mathematician. Our professional needs were reasonably fulfilled, I repeat, before the end of the previous Millennium. No one suspected however (but maybe for a couple of real visionaries) that the revolution put forward by a little bunch of theoretical physicists would be growing and growing without an end (it is still on its way today, faster than ever) to first equal, and subsequently surpass, all previous revolutions in the history of Humanity—as the steam machine, the railroad, the telegraph/telephone, or the cinema, or whichever you may right now think of—for its enormous and almost simultaneous impact at world scale that has made, in many respects, of our planet Earth a global village. Its projection to the future, still to come, will be for sure not less impacting. And it even seems now to be not so dificult to predict, as I will try to show in a forthcoming section.

---

[3] "Power tends to corrupt, and absolute power corrupts absolutely" as Lord Acton liked to observe. In the present context, we may say that PPT corrupts absolutely, in the sense that it doesn't have to be used always, but only when it is really needed [J. Martín Ramírez].

## 6. Once more the same discourse, but from a different perspective

This was, of course, already in the 1980's, an available possibility to send preprints abroad, and it was in fact being used in some specific cases; but it could not be adopted as a real solution to the problem we had (see the beginning of this Chapter). For many obvious reasons, starting with low bandwidth, long transmission times, and the very high associated costs. However, it was very much employed by the administration, big companies, etc. During a visit that I paid to the Physics Department of Leningrad University in 1989 (Gorbachev's perestroika time), in order to initiate scientific relations between our Physics Departments, this was my only reliable, fast daily connection to the outside world.

Actually, a predecessor of the Fax concept had been already invented in the 19th Century, although the first commercial version of the modern fax machine was patented in 1964 by Xerox Corporation and got the name of Long Distance Xerography (LDX). No doubt, xerox-copying and printing is a main ingredient in the process. By the late 1970's many companies (mainly Japanese ones) were in the fax market [Fax 2016]. The way the fax works, the fax technology, has improved considerably with the years. In particular, in the early 1980's Ethernet enabled fax services were already in operation. But, as for now, I just want to concentrate on the idea of the fax machine as a very clever way, more or less sophisticated, to transmit all kinds of printed material, that is both text and images.

Seen from this point of view, the very first steps of the internet revolution aimed at a very similar goal: to transmit on the internet line (in a way quite similar to the telephone line), the LATEX files corresponding to the scientific preprints we were interested in. In a way, there was not much difference of principle between these two conceptions, only technical issues and, of course, associated costs, as explained. The Fax concept has not substantially evolved since that time, although, to be fair, one must consider, separately on its own right, optical scanning [Optical-scanner 2016], one of the main ingredients of the fax machine, namely the encoding in a file of the information contained in the scanned picture—in a compressed way, always with some loss with respect to the original, depending on the nominal resolution of the device we are using. In fact, scanning has become a habitual task in our daily life now and, by the way, an important component of the whole edition process: the addition and integration of pictures into the final document, when our book or article in question contains images, and not just mathematical figures. Those image files could thus be added very easily to the LATEX document in the process of compilation of the corresponding file.

Anyway, the enormous possibilities, the final evolution of the brilliant idea of first encoding an article or book in a LATEX file, then transmitting this file from computer to computer through the internet, and finally reproducing the whole book or article by compiling the file—important to remark, without ever losing not the least single bit in the whole process!—had such an incredible potential that their uncountable applications could not be foreseen in the 1990's.

And the question is now, why was it so? What on Earth makes this concept, this very simple idea so extremely powerful and endlessly fruitful?

## 7. The uncountably many present uses of the Cyberspace

The key hint for an answer in two sentences. The initial solution found to the problem of preprint distribution was to reduce the content of the preprint to a code [Latex code 2016], without any information loss. The written pages, possibly including mathematical plots (`figures'), that is, all the information of the preprint (or book) pages was encoded in a file. After sending this file through the internet, the end computer got this file from which THE WHOLE content of the initial object was recovered. In fact, the initial object itself! or one completely indistinguishable from it, when the file was printed. And now here comes the revolutionary idea. Just looking around you, there are many different things that could be codified, at least in principle, e.g., their essence reduced to a (more or less long) file, which might be transmitted and then reproduced at a position at the other end of the World (given such a wide network would be available).

The first addition to the original issue was the incorporation of images to the text/plot files, in other words, image encoding [Image encoding 2016]. This was done by optical scanning, as already discussed. In this respect, the following very important consideration is here in order. The number of letters in any alphabet is always finite, even if you want to encode those of all languages written on Earth, and in all possible different fancy fonts that may have been invented (of course handwriting is a completely different story, see later), so that you can choose the one you like more as final result, for your purposes. All this gives, in the end, a finite number of possibilities, which can be easily encoded and stored in personal computers of an ever-growing capacity.

Function plots are already a bit different, since they rely on appropriate mathematical programs for the approximation of function plots (usually Mathematica [Mathematica 2016], Maple [Maple 2016], or the like), which should be involved in the play. They use their own techniques for approximating any given function to a desired accuracy [SAT 2016]; but this works again pretty well. Images, on the contrary, are a completely different affair.

How do you capture the essence of a picture? Certainly, this has been done many times, with more or less success and under different circumstances, by painters, and later with the help of photo cameras. Now the problem is to digitize this, to encode this fabulous amount of information under the form of a binary file, as in the original case of the preprint. The problem is of a truly profound mathematical nature: to convert a continuous many-dimensional variable into a discrete one, to codify using 1's and 0's a continuous quantity that has, in principle, infinitely many possibilities, as the infinitely many different colors of a Renoir painting at infinitely-many positions (dots) on a canvas. Now, this can be done, but the clear difference with the simple case of the information contained in an ordinary book or preprint (without pictures) is that with images the codification of its content is no more lossless. By approximating the infinitely many possibilities by a finite number of pixels you will always loose some information. A good codification procedure will have a minimal loss with an also minimal length (or `weight') of the resulting code or file. Optimal compression is a very essential concept that comes here into play and which, given the importance of this development, has quickly developed into a flourishing industry worldwide [Optimal compression 2012] .

But let us go back to our initial description and try to recapitulate. We scientists started by trying to improve our new inventions of TEX and the web by including a picture or two of some simple device or illustration, appearing in our preprint or booklet, and sending both together, always with pure scientific purposes. But immediately afterwards normal people realized the enormous power of this new toy; they bought themselves a computer and went on to send through the internet family pictures, and then movies (why not? that are nothing else but collections of pictures), and so on and so forth. And finally the computer was unnecessary and they started to do this all the time from anywhere with their small and very fancy mobiles. The initial, triangular thin path connecting the computers in Los Alamos, CERN and DESY evolved into an ever growing network of ways, roads, and highways. To end up becoming a truly global network, like that of a gigantic spider covering the whole Earth, which was given the most appropriate name one could ever have thought of: The World Wide Web (www or just Web) [W3 2016].

But this is of course not the end, rather the mere beginning. As the next possibility after sending pictures and movies of all kind you will immediately think of sending three-dimensional bodies. Any item will be reduced to small components, and with the advent of 3D printers [3dprinting 2016], after recovering the code of the device at destination you can 3D print it and obtain an instant copy of your favorite toy or working machine anywhere. And after you connect this development with no less important ones taking place in all other fields of scientific, humanistic, cultural, artistic or technological knowledge, the power of this `transmission idea' becomes overwhelming, unstoppable: we miss the words to describe it.

At the level of remote control of a device, when you do not transmit a physical object, but only an order to a remote machine, this procedure is already extremely helpful. Think, just of medicine: medical operations, the most delicate surgeries, diagnostic and control processes in medical praxis are now routinely carried out from the distance by the best of the specialists available. I am sure the reader, on his/her own, will find lots and lots of different examples to add here.

## 8. The dangers of the Cyberspace

But not everything is positive, with the Cyberspace. The dangers of this new, ubiquitous phenomenon are quite obvious, and have been stressed very often, e.g., by the Wall Street Journal. I copy from there:
> "Do we know what our kids are doing online? What do you think are the best ways to protect your kids in a world where the walls between online and online are coming down?" [Ilgenfritz 2008]

Sure, on general grounds the situation is not new at all. It has been repeated in human history every time a new invention has appeared; starting from the very first remarkable discoveries, as those of the wheel (known already in the Chalcolithic, 4,000 years BC, before the Early Bronze Age) [Wheel 2016], and the lever (known to Archimedes, 3rd century BC: "Give me a place to stand, and I shall move the Earth with it") [LeverLaw 2016], and continuing with much more recent crafts, as the airplane (flown for the first time by the Wright brothers in 1903) [Airplane 2016], and a lot many others. In spite of the generically good purposes of the authors of all these inventions, always aimed at easing human work and improving performance and possibilities, soon all those new instruments were used by the armies of all ages to develop evermore powerful deadly weapons.

In the case of the Cyberspace some of the dangers are much subtler, and most of them have nothing to do with the original purposes of the Arpanet, the American military network at the very origin of the internet. It is certainly true that among the most fearful dangers of the Cyberspace are its possible uses by all sort of international terrorists [Weimann 2015], organized criminals [Broadhurst 2013], mafia cartels [Agarwal 2006], drug dealers [Ngo 2014], and so on. But there are other very serious dangers of the internet that, having also to do with crimes, those are of a very different kind, as cyberbullying and harassment, in its various forms (leading too often to suicide), sexual predation targeting children in chat rooms, too easily accessible pornography, damaging person's reputation [Kam 2007], and several other. All those issues have been suficiently addressed, and in much depth, in the references just given above and elsewhere, also feasibly in other chapters of the present book. I will do not deal with them in any more detail.

Instead, I want here discuss, in a little bit more extent, some rather subtle point that, however, cannot escape anybody's attention. Everywhere, everyday, while just walking, or traveling to work, or sitting on a bench in a park for a short rest, eating at a restaurant, summing up, anywhere, anytime, you will be surrounded by people of all ages and condition attentively looking at a small device, sometimes (but not that often) talking to it, while they hold it on one hand, namely a fancy cellphone. An increasingly large number of people heavily depend on this device: their personal connection to the Cyberspace. Some of them have become absolutely fascinated by it, they have even got addict to it, and cannot stop looking at their cellular screen constantly. In a word, we could say that these people actually live in the Cyberworld, rather than in the real one.

We may certainly view, at least some of them, as actual slaves of the Cyberspace: a new, contemporary form of slavery [Blume 2015] where individuals are not bound any more by ropes or heavy chains, against their will, but where they freely choose to be bound to this web of very thin, invisible, but extremely strong threads (for all we may judge): the internet. I will not go into the problem in much further detail, in the first place because it is none of my competences. But I did want to rise the point because it is a very serious one that really worries me and cannot go unnoticed. As I said before, it was no surprise, on the contrary, it actually was to be expected, from previous experiences, that the military on one side and the criminal forces, on the other, would immediately try to make use of the new invention at discussion here, as it has always been the case in past history. But this other issue I am addressing now is a brand new development, a kind of new illness without any precedent with past inventions. Possibly it will just be a strong but passing fever but this is not all that clear right now.

Of course one might argue that, in principle, there is nothing wrong with making your free choice to join the Cyberspace. Everybody can make extraordinarily good use of the immense possibilities it offers, as conveniently stressed in the preceding sections—and many people do so, in fact, even a certain percentage of those internet addicts, I would dare say. But what is really worrisome is the huge amount of people, young and old, who, quite on the contrary, are too much distracted by the internet with very trivial (if not dangerous, as seen above) issues, which rob them of all their precious time. So that they are left with no single minute to carry out any positive task: learn, work, concentrate on some important issue, etc. Maybe this formulation is taking things to the extreme, but anyhow it is not so far from truth, in regretfully too many cases.

## 9. The future of the Cyberspace

I do not dare to make a list, not even a sketch of the extraordinary number of future capabilities of the internet concept [Breene 2016]. The advances in genetics and medicine, with the discovery of the genetic code of animals and humans [Radford 2003], the creation of human organs from cells (now not necessarily mother cells) [Weisberger 2016], the physical and technological exploration of the nano, pico and femto scales [Bradley 1997], ... All these investigations rely on mathematical modeling, on the reduction of a real object or device to a certain file, which encodes all relevant information of the original (sometimes absolutely all of it). And once you are there, you can then send the file anywhere on our Planet or even on tiny little ships, the size of a postcard or less, to visit different stars and extraterrestrial planets, in a universal journey soon to be undertaken by the human society [Feltman 2016].

With the use of quantum technology, the possibilities to encrypt information [Preskill 2016] in a secure way and those of teletransporting it to far distances could make it possible, in a not so far away future, to send egg cells or nanocomponents, or pico-machines and devices (still to be conceived), with the aim to colonize our Universe. And this would happen in a very different way to the one imagined until very recently, which always involved big, manned, shining spaceships. However, one should be very careful with the quantum world [Merali 2015]. In fact, in this case the result is a true teletransportation, since the very important no-cloning theorem [No-cloning 2016] prevents the cloning by the Einstein-Podolsky-Rosen procedure [Fine 2013] of any quantum state. In fact, the brand new state emerges, far away, only and at the very same instant when the original state disappears, ceases to exist [No-cloning 2016].

The concepts of artificial intelligence and of artificial life will be further exploited, with consequences that are dificult to assess in a rigorous way yet. Should we be afraid of robots? And by this I mean, of replicants, of perfectly working copies of a human being, capable to pass all intelligence tests [Veselov 2014] available; what will make them indistinguishable from a real person. This has already appeared, and quite often, in science fiction movies. But what I here mean is that, at some point, this issue will no more be fiction science, but true, well established science.

Will we humans be able to live for ever (or at least for a very, very long time) in one of these new bodies? —our self having been captured, as a whole, in a sort of extended code and then having been put inside the replica, right here or millions of miles away. Will our soul be eventually replicated, also? And that, with all our memories, and with all of our feelings, as love, and fear, and freedom, and anger, too?

## 10. Conclusion

We started our project by aiming at the codification of a mere set of letters, of words and sentences—of more or less important scientific results—and at sending them `for free' to distant places, and have now ended up by pretending to codify our own body and soul. So powerful is the Cyberspace concept that resulted as the end product from the process. An extremely simple, but enormously clever, solution to a very down-to-earth problem, just aimed at saving time and money, was the tiny seed that grew up unsteadily, to become a monster network of ever growing, immense possibilities.

**Acknowledgement**. The author was partially supported by MINECO (Spain), Project FIS2013-44881-P, by CSIC, I-LINK1019 Project, and by the CPAN Consolider Ingenio Project.

[Fax 2016] https://en.wikipedia.org/wiki/Fax

[Faxauthority 2016] https://faxauthority.com/fax-history/

[Feltman 2016] Rachel Feltman, Stephen Hawking wants to use lasers to propel a tiny spaceship to Alpha Centauri (The Washington Post, 2016), https://www.washingtonpost.com/news/speaking-of-science/wp/2016/04/12/stephen-hawking-wants-to-use-lasers-to-propel-a-tiny-spaceship-to-alpha-centauri/

[Fine 2013] Arthur Fine, The Einstein-Podolsky-Rosen Argument in Quantum Theory (Stanford Encyclopedia of Philosophy, 2013), http://plato.stanford.edu/entries/qt-epr/

[Ginsparg 2016] Paul Ginsparg, personal webpage, http://infosci.cornell.edu/faculty/paul-ginsparg

[Http 2016] https://en.wikipedia.org/wiki/Hypertext Transfer Protocol

[Ilgenfritz 2008] Stephanie Ilgenfritz, The Real Dangers of Cyberspace (The Wall Street Journal, 2008), http://blogs.wsj.com/juggle/2008/08/22/bullies-and-instant-messagethe-real-dangers-of-cyberspace/

[Image encoding 2016] Image encoding (Code.org, 2016), https://code.org/files/CSPUnit1Lesson6.pdf

[Infotheor 2016] https://en.wikipedia.org/wiki/Information_theory

[Internet 2016] https://en.wikipedia.org/wiki/Internet

[Kam 2007] Katherine Kam, 4 Dangers of the Internet (WebMD, 2007), http://www.webmd.com/parenting/features/4-dangers-internet#5

[Knight 2014] Dan Knight, Personal Computer History: The First 25 Years (Low End Mac, 2014), http://lowendmac.com/2014/personal-computer-history-the-first-25-years/; https://en.wikipedia.org/wiki/History_of_personal_computers

[Knuth 1969] Donald Knuth, The Art of Computer (TAOCP) (Addison-Wesley Pub. Co., 1969); http://www-cs-faculty.stanford.edu/knuth/brochure.pdf

[Lamport 1986] Leslie Lamport, "LATEX: A Document Preparation System" (Addison-Wesley, 1986).

[Lanl 2016] http://www.lanl.gov/

[Lariviere et al 2013] Vincent Larivière, Cassidy R. Sugimoto, Benoit Macaluso, Staša Milojević, Blaise Cronin, and Mike Thelwall, ArXiv e-prints and the journal of record: An analysis of roles and relationships (ArXiv, 2013), https://arxiv.org/ftp/arxiv/papers/1306/1306.3261.pdf

[Latex code 2016] https://en.wikibooks.org/wiki/LATEX/Source_Code_Listings

[LeverLaw 2016] https://www.math.nyu.edu/ crorres/Archimedes/Lever/LeverLaw.html

[Lewin 2016] Walter Lewin, MIT Lectures (2016), https://www.youtube.com/playlist?list=PLyQSN7X0ro203puVhQsmCj9qhlFQ-As8e

## 12. Author's short biosketch

Prof. Dr. Prof. H.C. Emilio Elizalde was born in Balaguer (Spain) in 1950. He obtained his MS in Physics, MS in Mathematics, and PhD in Physics from Barcelona University, first and last with Extraordinary Master and Doctorate Awards. Elizalde was Humboldt Fellow in Hamburg and Berlin (Germany), and SEP Fellow in Japan. He held visiting Scholar/Research Contracts at Hamburg U, the MIT, the CfA at Harvard U, PennState U, Trondheim NTNU, the KEK, Hiroshima U, Jena U, Leipzig U, the NTZ, and Trento

U. Professor of Mathematics and Physics at Barcelona University for over twenty years. Since 1993 he is a member of the National Higher Research Council of Spain (CSIC), where he is a Senior Research Professor and Principal Investigator of various projects—including a top rated Consolider one (group project)—and executive board member of both a Consolider-Ingenio 2010 (CPAN) and a European Union project (CASIMIR). Elizalde is the founding leader of the "Theoretical Physics and Cosmology" Group of ICE-CSIC and IEEC, highly recognized internationally, unchallenged in Spain in its field, and leading the ranking in Normalized Impact Factor of SCIMAGO World Reports 2009-13. He got an Honorary Professorship from Tomsk TSPU University, Russia, and is recipient of the Gold Medal of TSPU. Elizalde has published groundbreaking works on zeta functions, the Chowla-Selberg formula, and cosmology. One zeta function is named after him. He is very proud of his former students, a number of them being highly reputed scientists now.